\documentclass[aps,prb, amsmath,amssymb,floatfix,12pt, longbibliography]{revtex4-1}
\usepackage{tabularx}
\usepackage{bm}
\usepackage{euscript}
\usepackage{graphicx}
\usepackage{color}
\usepackage{amsfonts}
\usepackage{exscale}
\usepackage{amsbsy}
\usepackage{subfigure}
\usepackage{textcomp}
\usepackage{comment}
\usepackage{refcount}

\usepackage{hyperref}

\numberwithin{equation}{section}

\newcommand{\be}{\begin{equation}}
\newcommand{\ee}{\end{equation}}
\newcommand{\bea}{\begin{eqnarray}}
\newcommand{\eea}{\end{eqnarray}}
\newcommand{\tr}{\textrm{tr}}

\renewcommand{\t}{\tilde}

\renewcommand{\l}{\lambda}

\begin{document}

\title{Non-Fermi liquid Superconductivity: \\ Eliashberg versus the Renormalization Group}
\author{Huajia Wang$^{\Delta}$, Srinivas Raghu$^{{\bar \psi},\psi}$, Gonzalo Torroba$^{\phi}$}
\affiliation{$^\Delta$ Department of Physics, University of Illinois, Urbana IL, USA}
\affiliation{$^{\bar \psi}$Stanford Institute for Theoretical Physics, Stanford University, Stanford, CA 94305, USA}
\affiliation{$^\psi$SLAC National Accelerator Laboratory, Menlo Park, CA 94025, USA}
\affiliation{$^\phi$Centro At\'omico Bariloche and CONICET, Bariloche, Rio Negro R8402AGP,
Argentina}
\date{\today}

\begin{abstract}
We address the problem of superconductivity for non-Fermi liquids using two commonly adopted, yet apparently distinct methods: 1) the renormalization group (RG) and 2) Eliashberg theory.  The extent to which both methods yield consistent solutions for the low energy behavior of quantum metals has remained unclear.  We show that the perturbative RG beta function for the 4-Fermi coupling can be explicitly derived from the linearized Eliashberg equations, under the assumption that quantum corrections are approximately local across energy scales.  We apply our analysis to the test case of phonon mediated superconductivity and show the consistency of both the Eliashberg and RG treatments.  We next study superconductivity near a class of quantum critical points  and find a transition between superconductivity and a ``naked" metallic quantum critical point with {\it finite}, critical BCS couplings.  
We speculate on the  applications of our theory  to the phenomenology of unconventional metals.  
\end{abstract}

\maketitle

\tableofcontents

\section{Introduction}\label{sec:intro}

Many strongly correlated electronic systems exhibit surprising non-Fermi liquid behavior, often experimentally connected with an enhancement in the superconducting critical temperature $T_c$, incoherent pairing, and proximity to a quantum critical point\cite{Stewart1984, Schofield1999, Stewart2001, Coleman2005, Lohneysen2007,Gegenwart2008, Dressel2011, Shibauchi2013, Kuo2016}. However, it has been difficult to construct a fully controlled framework to explain these effects.
A central problem in condensed matter physics is then to understand the interplay between non-Fermi liquid effects and superconductivity.

Considerable analytical\cite{Holstein, Hertz1976, Millis1988, Reizer1989, Littlewood1992, Millis1993, Polchinski1994,Altshuler1994,  Nayak1994a, Chakravarty1995, Son:1998uk,Oganesyan2001, Varma2002, Abanov2003, Metzner2003, Chubukov2005, Rech2006, Lawler2006, Fradkin2007,Lee2008, shankar2008, Zaanen2008, Lee2009, Metlitski2010, Moon2010, Mross2010, Yamamoto2010,semi-holographic, Chung2013, Mahajan2013, Phillips2013,Lee2013, Fitzpatrickone, FKKRtwo, Sur2014, Fitzpatrick:2014efa, Torroba:2014gqa, Fitzpatrick:2014cfa, ZWang2014, Metlitski, Raghu:2015sna, Meszena2016, 2016arXiv160601252W, PhysRevB.91.064507, Mandal:2016abm} and numerical\cite{Berg2012, Schattner2016, Wang2016} efforts have been devoted to the study of this subject.  Nonetheless, key aspects of the problem of non-Fermi liquid behavior at quantum critical points and the interplay with superconductivity remain poorly understood.   Therefore, it is crucial to strengthen  existing analytical approaches, and to develop new ones. The main analytical tools in this area are the Eliashberg equations (a strong coupling version of gap equations) and the renormalization group (RG). Both approaches look in principle quite different. The Eliashberg theory gives a set of coupled equations that need to be solved for the gap and wavefunction factors. These equations mix modes ranging from high energy all the way to those at the Fermi surface, and hence could be sensitive to mixing of the UV and IR degrees of freedom. The RG, in contrast, is based on the Wilsonian effective action and encodes the superconducting instability as dimensional transmutation in the BCS 4-Fermi coupling\footnote{This simply means that an emergent scale, namely that of the superconducting gap or T$_c$, arises at low energies out of a classically marginal coupling --in this case, the 4-Fermi BCS coupling.}. The general relation between both frameworks has not been elucidated so far, and it is not clear when they can yield equivalent results. 

For Fermi liquids, Bardeen, Cooper and Schrieffer (BCS)\cite{Bardeen:1957mv} established how superconductivity occurs by solving the gap equation for quasiparticles. The value of the physical gap was later understood within the RG approach of Shankar\cite{Shankar} and Polchinski\cite{Polchinski}, as dimensional  transmutation of the 4-Fermi BCS coupling. For non-Fermi liquids, however, the situation is much more challenging and not fully understood. This is in part connected with the appearance of new bosonic modes, such as the critical order parameter fields. They can enhance the effective attraction between fermion pairs, but also lead to a stronger quasiparticle decay rate (due to anomalous wavefunction and mass renormalization of the fermionic quasiparticles). Since the later tend to decohere the Fermi surface, non-Fermi liquids can exhibit a competition between pair-enhancing and breaking effects, leading to a complicated dynamics where it is not clear which effect prevails. At the same time, this presents an extremely interesting opportunity to explain some of the observed properties of unconventional metals and their underlying quantum critical behavior.

Our approach in previous works\cite{Torroba:2014gqa, Fitzpatrick:2014cfa, Fitzpatrick:2014efa, Raghu:2015sna} has been to develop a consistent renormalization group description for non-Fermi liquids with critical bosons.
 It was found in\cite{Fitzpatrick:2014cfa, Raghu:2015sna} that the quasiparticle wavefunction and mass renormalizations lead to a new term in the BCS beta function, linear in the 4-Fermi coupling and proportional to the anomalous dimension. This term has important physical consequences\cite{Raghu:2015sna}: it allows to interpolate between regimes of coherent and incoherent superconductivity (in this case the non-Fermi liquid energy scale is bigger than the superconducting gap), and moreover it gives rise to a novel quantum critical state with finite BCS coupling. However, studies of the Eliashberg equation so far have not found the analog of this mechanism; in fact it has been recently suggested in\cite{2016arXiv160601252W} that the gap equation for non-Fermi liquids could contain nonlocal effects that are not captured by the RG. This situation has cast doubt on the validity of the RG and its connection with the Eliashberg formalism. 
 
The goal of this work is to explain the connection between the Eliashberg and RG formalisms for non-Fermi liquids. While our immediate motivation was to clarify the above situation in models with overdamped critical bosons (such as those that appear near symmetry breaking transitions with zero momentum), the approach we will develop is much more general. It can be applied to any system where the formation of superconductivity is described by the Eliashberg equations. A large variety of models fall into this category, including phonon-mediated superconductivity, ferromagnetic and anti-ferromagnetic quantum critical points, and so on.
Our main general result will be a derivation of the RG beta functions as an approximation to the Eliashberg equations when a condition of energy locality is satisfied. This will provide a unified RG treatment for a large class of models, revealing the dynamics of their 4-Fermi coupling and its relation with the gap function. The condition of energy locality is checked to be satisfied in specific examples of interest, such as phonon-mediated superconductivity and non-Fermi liquids with overdamped bosons. This shows explicitly the agreement between the Eliashberg and RG predictions

Let us stress before  we start that our analysis pertains to the onset of the superconducting instability --it is done at the linearized level for the gap and at zero temperature. In the future, it will be important to perform a full finite temperature study (see \cite{2016arXiv160601252W} and\cite{GTtalk}), as well as a zero temperature analysis including nonlinear effects from the gap.

The work is organized as follows. In Sec. \ref{sec:RG}, we present the general analysis underlying the equivalence of the Eliashberg equations and the perturbative RG flows, and state the conditions required for this equivalence.  In Sec. \ref{sec:elph}, we apply our analysis to the test case of the electron-phonon problem; besides reproducing the known results for the superconducting case, we derive a beta function for the 4-Fermi coupling that encodes quantum effects from high frequencies. In Secs. \ref{sec:NFL} and \ref{sec:interplay}, we apply our analysis to superconductivity near a class of quantum critical points. We study the transition between the normal and superconducting regimes both numerically and analytically, finding a BKT scaling that is consistent with the fixed point annihilation picture from the RG\cite{Raghu:2015sna}. We also discuss applications of the present results to unconventional metals. We end in Sec. \ref{sec:conclusion} with our conclusions and some future directions.

\section{RG from the Eliashberg equations}\label{sec:RG}

In this section we present our general approach for deriving the RG from the Eliashberg equations. We consider nonrelativistic fermions at finite density, interacting with an arbitrary 4-Fermi coupling $u(p_0)$ that depends on frequency. Physically, this interaction arises form integrating out bosonic modes. Under a key assumption of energy locality, which we will discuss in detail, we will obtain the beta function for the 4-Fermi BCS coupling, and show that it correctly reproduces the quantum corrections. 

This has various advantages. First, as discussed in the Introduction, many different models of strongly correlated electrons fall into the general Eliashberg equations that we analyze. Our approach gives a unified treatment for all these theories, and provides additional physical insights into their dynamics. As a canonical example, we will work out the beta function for the electron-phonon problem. Furthermore, obtaining the beta function from the Eliashberg equation puts the RG approach on a firmer footing, since it does not require additional assumptions for the scalings of the different fields.

\subsection{Schwinger-Dyson-Eliashberg equations}\label{subsec:eliashberg}

In this work we will focus on the class of finite density electronic systems that can be described by the Eliashberg equations
\bea\label{eq:El1}
\left(Z(p_0)-1\right)p_0&=& \frac{1}{2}\int dq_0\, u(p_0-q_0)\,\frac{q_0}{\sqrt{q_0^2+|\Delta(q_0)|^2}} \,,\nonumber\\
Z(p_0)\Delta(p_0)&=&\frac{1}{2N}\int dq_0\, u(p_0-q_0)\,\frac{\Delta(q_0)}{\sqrt{q_0^2+|\Delta(q_0)|^2}}\,.
\eea
As we discuss shortly, these arise in QFT as a truncation of the Schwinger-Dyson equations, so we will often refer to them as the Schwinger-Dyson-Eliashberg (SDE) equations.

The parameters $Z(p_0)$ and $\Delta(p_0)$ correspond to the wavefunction renormalization and gap function of the finite density fermions, namely they appear in the following terms in the fermionic Lagrangian:
\be
L_f=- \psi(p)^\dag(iZ(p_0)p_0-\varepsilon_p) \psi(p)+ Z(p_0) \Delta(p_0) \psi(p) \psi(-p)+\text{h.c.}
\ee
where $\varepsilon_p$ is the quasiparticle energy.
The kernel $u(p_0)$ will be kept as an arbitrary (even) function in the general analysis of this section, and in later sections we will specialize it to different systems. The reader may be puzzled at the fact that the two integral equations in (\ref{eq:El1}) have different prefactors.  At this stage, $N$ is some adjustable parameter that depends on the relative strength of the corrections to $Z$ and $\Delta$.  In Sec. \ref{sec:NFL}, we show how this comes about in a specific example.

\begin{figure}[h!]
\centering
\includegraphics[width=1.3\textwidth]{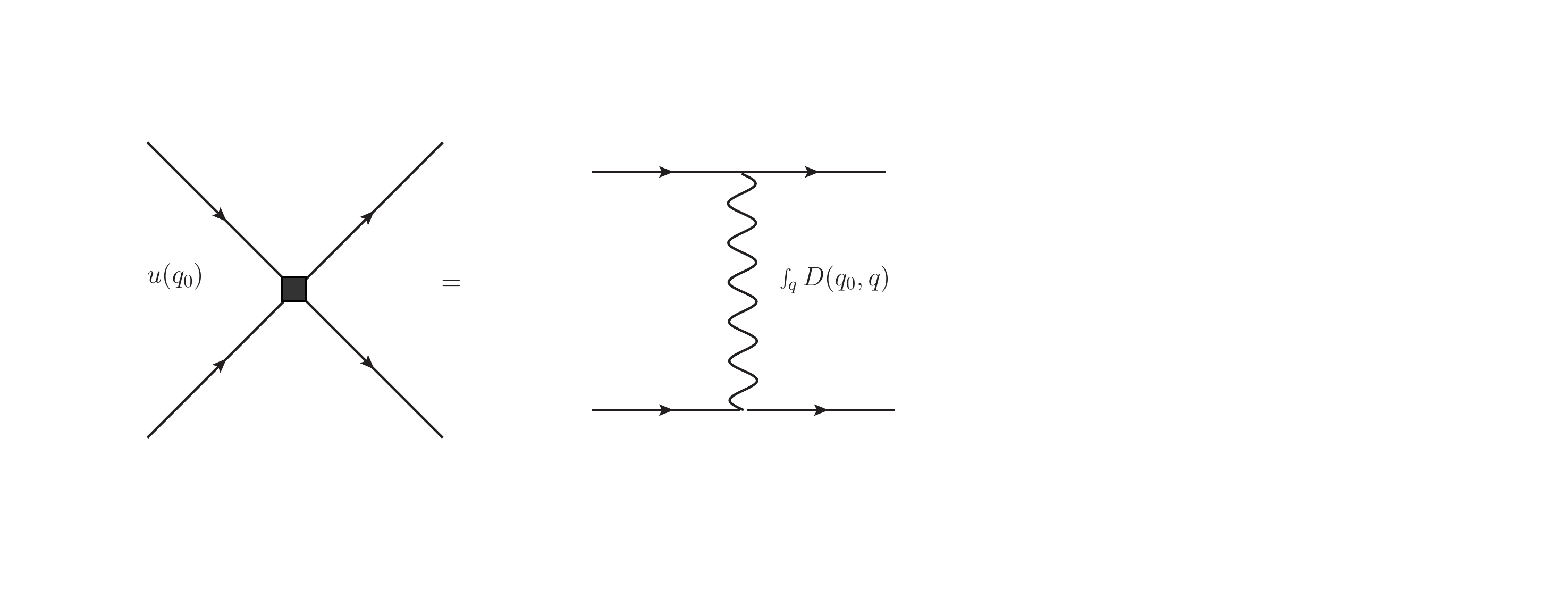}
\caption{Frequency-dependent kernel $u(q_0)$ generated by boson exchange.}\label{fig:diagrams1}
\end{figure}

\begin{figure}[h!]
\centering
\includegraphics[width=1.4\textwidth]{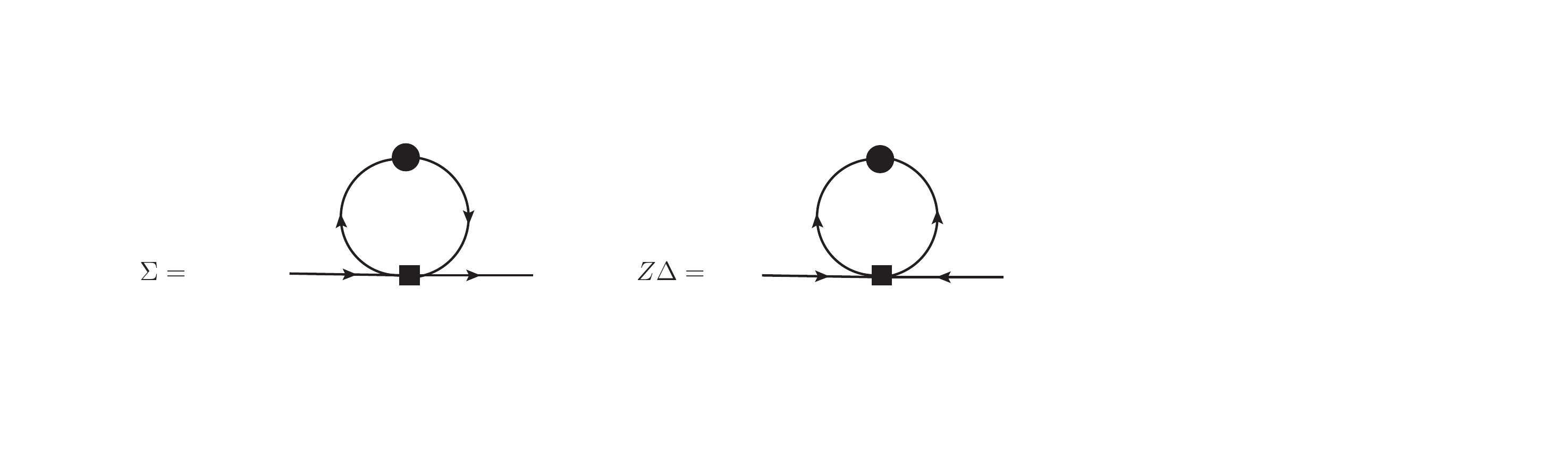}
\caption{Diagrammatic representation of the Eliashberg equations.}\label{fig:diagrams2}
\end{figure}

The dynamics encoded by the SDE equations arises from coupling fermionic quasiparticles to a bosonic mode, with Green's function denoted by $D(q_0, q)$. A key requirement for the validity of Eliashberg theory is that the typical boson mode propagates much more slowly than the fermion. In addition to suppressing vertex corrections (Migdal's theorem), this kinematic requirement ensures that the boson scatters fermions primarily in the direction tangential to the Fermi surface.  If this kinematic constraint is satisfied, we may integrate the tree-level exchange of bosons over the Fermi surface, which leads to a frequency-dependent kernel, that we have called $u(q_0)$ above,
\be\label{eq:uD}
u(q_0) = \int_{q_\parallel}\, D(q_0, q)\,.
\ee
This is shown in Fig. \ref{fig:diagrams1}. The boson propagator is taken to be even under $q_0 \to -q_0$, so that $u(q_0)$ is also even. Given this, the Eliashberg equations are the Schwinger-Dyson diagrams shown in Fig. \ref{fig:diagrams2}. We also recall that $Z$ is related to the quasiparticle self-energy $\Sigma$ by
\be\label{eq:Sigmadef}
Z(p_0) p_0 = p_0 + \Sigma(p_0)\,.
\ee
In order to derive this closed set of coupled equations, it was crucial to neglect vertex corrections. We will present below some concrete examples of the underlying calculations.

These are the nonlinear SDE equations, and we will perform an additional approximation by linearizing in the gap. The intuition is that if a physical gap develops at low energies, i.e. $\Delta(p_0=0) \neq 0$, the system becomes gapped at energy scales below $\Delta(0)$, and the nontrivial frequency dependence stops. On the other hand, at energies much larger than the physical gap, which we denote by $\Delta_*$, the effects of this mass become negligible. Therefore, we can linearize the right hand side of (\ref{eq:El1}) while introducing an IR cutoff of order $\Delta_*$,
\bea\label{eq:El2}
\left(Z(p_0)-1\right)p_0&\approx & \frac{1}{2}\int_{|q_0|>\Delta_*} dq_0\, u(p_0-q_0)\,\text{sgn}(q_0)\,,\nonumber\\
Z(p_0)\Delta(p_0)&\approx &\frac{1}{2N}\int_{|q_0|>\Delta_*} dq_0\, u(p_0-q_0)\,\frac{\Delta(q_0)}{|q_0|}\,.
\eea
Note that the equation for the self-energy has decoupled from the gap function (except for the dependence on the IR cutoff), and so can be solved directly. This means that the one loop result is exact within the Eliashberg framework, because there is no wavefunction renormalization dependence on the right hand side of the first equation in (\ref{eq:El2}). Once $Z(p_0)$ is calculated in this way, the second equation becomes an integral equation that needs to be solved self-consistently for the gap function.

A different interpretation that yields the same result is to imagine working at finite temperature, near the critical temperature $T_c$ at which the gap vanishes. It is then sufficient to consider the linearized approximation. If the temperature is treated as an IR cutoff, the gap equation agrees with (\ref{eq:El2}), with $T_c = \Delta_*$. There are situations where this approximation can fail --for instance with first order transitions-- and then a full nonlinear analysis is necessary. At finite temperature, the discreteness of the Matsubara frequencies can also introduce new effects\cite{2016arXiv160601252W,GTtalk}. In this work, however, we restrict to the linear approximation and work at zero temperature, postponing a complete analysis to the future.

\subsection{The local approximation}\label{subsec:local}

Our goal is to derive a Wilsonian RG for the previous systems. 1PI calculations, such as those in (\ref{eq:El2}), involve mixing of high and low frequencies; in contrast, the RG approximation is essentially local in energy scale, because it arises from integrating out infinitesimal shells of momentum modes. In order to recast the Eliashberg equations in RG language, we then need to approximate the UV/IR frequency mixing by a local process. This is done by splitting the integration range into $|q_0|<|p_0|$ and $|q_0|>|p_0|$, and Taylor expanding the boson kernel in each range:
\bea\label{eq:local1}
u(p_0-q_0) &\approx & u(p_0) - q_0 u'(p_0) +\ldots\;,\;\text{for}\;\;|q_0|<|p_0| \nonumber\\
u(p_0-q_0) &\approx & u(q_0) - p_0 u'(q_0) +\ldots\;,\;\text{for}\;\;|q_0|>|p_0|\,.
\eea
(Recall that $u(p_0)$ is an even function). If $u(p_0-q_0)$ diverges at $p_0=q_0$, the singularity needs to be integrable so that the above splitting be valid. In this way, the nonlocal retardation effects due to boson exchange are approximated by local contributions. We will see shortly how this leads to an RG description of the Eliashberg equations.

The local approximation can be improved systematically by including higher order terms in the Taylor expansion. A necessary condition for the approximation to be valid is that there exists some small parameter such that
\be\label{eq:local-cond}
|p_0^n u^{(n)}(p_0)| \ll |u(p_0)|\,.
\ee
Let us consider two examples that will be relevant below. For
\be\label{eq:uphonon}
u(p_0) = \frac{g^2}{p_0^2+M^2}\,,
\ee
where $g$ is some coupling and $M$ a high energy scale, the condition (\ref{eq:local-cond}) becomes $|p_0| \ll M$, and so it amounts to a low-energy expansion. This kernel appears in the electron-phonon model. A different class of kernels arises in quantum critical theories, where $u(p_0)$ has a power-law dependence,
\be\label{eq:upower}
u(p_0) = \frac{g^2}{|p_0|^{2\gamma}}\,,
\ee
in terms of the dimensionless constant $\gamma$ --the fermion anomalous dimension, as we review below. Then (\ref{eq:local-cond}) requires $\gamma \ll 1$, and so restricts us to the regime of small anomalous dimension. This is a familiar restriction of the perturbative RG: it is useful near critical dimensions where the relevant couplings become almost marginal and hence are naturally small.

Let us implement this approximation in (\ref{eq:El2}), beginning with the simpler equation for the self-energy. Splitting the integral into $|q_0|<|p_0|$ and $|q_0|>|p_0|$, and performing the expansion (\ref{eq:local1}) gives
\be
\Sigma(p_0) =- \int_{\Delta_*}^{p_0} dq_0 \left(u'(p_0) q_0 + \ldots \right)-\int_{p_0}^\infty dq_0 \left(u'(q_0) p_0+\ldots \right)
\ee
where the dots are higher order Taylor terms. Note that the leading dependence with zero derivatives cancels out since it gives an odd integrand. Both integrals can now be done explicitly, obtaining
\be
\Sigma(p_0)=- \left(\frac{1}{2}u'(p_0) (p_0^2-\Delta_*^2)+\ldots \right)+\left(p_0 u(p_0)+\ldots \right)\,,
\ee
from the small frequency and large frequency ranges, respectively. A constant term from $u(\infty)$ can be absorbed into a field redefinition, and is not shown. Given the condition (\ref{eq:local-cond}), the dominant contribution results from the range $|q_0|>|p_0|$,
\be\label{eq:Sigmalocal}
\Sigma(p_0) \approx p_0 u(p_0)\,.
\ee
This has all the expected properties for a Wilsonian self-energy: it is dominated by high frequency modes, and its sign is proportional to the external frequency.

For the gap equation, we split $\Delta(q_0)$ into even and odd parts; we keep only the even part since the odd one couples to derivatives of $u(q_0)$ and not to $u(q_0)$ itself, and hence is subleading. For the even part, we can restrict the integral to positive frequencies, and the local approximation becomes
\be\label{eq:local2}
Z(p_0) \Delta(p_0)=\frac{1}{N} \int_{\Delta_*}^{p_0} dq_0 \left(u(p_0)+\ldots \right)\frac{\Delta(q_0)}{q_0}+\frac{1}{N} \int_{p_0}^{\Lambda_0} dq_0 \left(u(q_0)+\ldots \right)\frac{\Delta(q_0)}{q_0}\,.
\ee
For future use in numerical calculations, we have introduced a UV cutoff $\Lambda_0$ for the frequency integral, which at the end can be taken to infinity. This gives an integral equation for the gap in terms of local in energy contributions. To make this more explicit, we now present an equivalent differential equation for the gap. A similar derivation appeared in\cite{Son:1998uk} for color superconductivity.

We will keep only the leading term of the Taylor series in (\ref{eq:local2}); in general, the validity of this has to be checked numerically, once the gap function has been determined. We will do this explicitly below for the electron-phonon system and for non-Fermi liquids. Let us also introduce the notation
\be
\t \Delta(p_0) \equiv Z(p_0) \Delta(p_0)\,.
\ee
Deriving once with respect to the external frequency $p_0$ gives
\be\label{eq:Deltap}
\t \Delta'(p_0) = \frac{1}{N} u'(p_0) \int_{\Delta_*}^{p_0}\, dq_0\,\frac{\t \Delta(q_0)}{Z(q_0) q_0}\,.
\ee
Next, we divide both sides by $u'(p_0)$ and derive once more, obtaining the differential equation
\be\label{eq:diff-gap-eq}
\frac{d}{dp_0} \left(\frac{\t \Delta'(p_0)}{u'(p_0)} \right)-\frac{1}{N}\,\frac{\t \Delta(p_0)}{Z(p_0) p_0}=0\,.
\ee
The last step is to find the appropriate boundary conditions.
The IR boundary condition is obtained from (\ref{eq:Deltap}):
\be\label{eq:physical-gap}
\t \Delta'(p_0=\Delta_*)=0\,.
\ee
For the UV boundary condition, we need an expression containing an integral between $p_0$ and $\Lambda_0$. For this, divide (\ref{eq:local2}) by $u(p_0)$ and then derive with respect to the external frequency:
\be
\frac{d}{dp_0}\left(\frac{\t \Delta(p_0)}{u(p_0)} \right)=-\frac{1}{N}\,\frac{u'(p_0)}{u(p_0)^2} \int_{p_0}^{\Lambda_0}\,dq_0\,u(q_0)\,\frac{\t \Delta(q_0)}{Z(q_0) q_0}\,.
\ee
The UV boundary condition is then
\be\label{eq:UVbc}
\frac{d}{dp_0}\left(\frac{\t \Delta(p_0)}{u(p_0)} \right)\Big|_{p_0=\Lambda_0}=0\,.
\ee

Since in the linearized approximation the overall scale of $\t \Delta(p_0)$ is not physical, only the ratio of the two integration constants of (\ref{eq:diff-gap-eq}) needs to be fixed. This is determined by the UV boundary condition. Having fixed the integration constant in this way, the physical gap $\Delta_*$ corresponds to the largest scale at which $\t \Delta'(p_0=\Delta_*)=0$. The fact that the physical gap is attained with vanishing derivative is consistent with the intuition above that the frequency evolution stops below $\Delta_*$.

\subsection{RG beta functions}\label{subsec:RGderivation}

We are now set to derive the RG beta functions from the Eliashberg equations.

The first RG beta function that we need is the anomalous dimension, related to the wavefunction factor $Z$ by
\be
2\gamma(p_0) = -  \frac{d \log Z(p_0)}{d \log p_0}\,.
\ee
The wavefunction renormalization can be calculated in closed-form from the first equation in (\ref{eq:El2}). If we further consider the local approximation, and assume that at strong coupling and low frequencies the self-energy $\Sigma(p_0)$ dominates over the tree-level kinetic term (this is usually the case in many systems of interest), then we have
\be\label{eq:gammalocal}
2\gamma(p_0) \approx  -  \frac{d \log u(p_0)}{d \log p_0}\,.
\ee
To derive this result we used (\ref{eq:Sigmadef}) and (\ref{eq:Sigmalocal}). As a simple example, for the kernel (\ref{eq:upower}), characteristic of quantum critical points, formula (\ref{eq:gammalocal}) correctly identifies the fermion anomalous dimension with the power $2\gamma$ in $u(p_0)$.

As we explained before, in the Eliashberg approach the bosonic modes are integrated out to produce a frequency-dependent 4-Fermi coupling; see e.g. Fig. \ref{fig:diagrams1}. We now need to compute the RG beta function for this coupling, which we denote by $\lambda(p_0)$. We take the convention that $\lambda>0$ corresponds to an attractive interaction along the BCS channel. 

The main property that the beta function should satisfy is that the superconducting instability appears as dimensional transmutation of the 4-Fermi coupling. It is useful to first recall how this works out in Fermi-liquid theory\cite{Shankar, Polchinski}, where the beta function is given by
\be\label{eq:BCSlambda}
p_0 \frac{d\lambda}{dp_0}=- \frac{1}{2\pi^2 N} \lambda^2\,.
\ee
The normalization --chosen for future convenience to be $1/(2\pi^2 N)$--
is model-dependent. Solving the differential equation shows that $\lambda \to \infty$ at an RG invariant scale
\be
\Lambda_* \approx \Lambda_0 e^{- 2\pi^2 N/\lambda_0}\,.
\ee
with $\Lambda_0$ a UV cutoff and $\lambda_0$ the value of the coupling at that scale. Since the attractive coupling is becoming very large, this signals the onset of the superconducting instability towards condensation of Cooper pairs; in this simple case, $\Lambda_*$ agrees with the physical gap $\Delta_*$ up to a prefactor.

We seek the analog of (\ref{eq:BCSlambda}) but for the theory described by the more general SDE equations discussed before. We want to retain the basic feature $\Delta \propto e^{-const/\lambda}$ of dimensional transmutation; furthermore, in the local approximation, the physical gap is determined by the condition (\ref{eq:physical-gap}). This motivates us to look for a 4-Fermi coupling of the form
\be
\lambda(p_0) = f(p_0) \frac{\t \Delta(p_0)}{\tilde \Delta'(p_0)}\,,
\ee
with $f(p_0)$ a function that will be fixed in order to yield a sensible beta function. We will now show that this functional form is required in order to obtain the right beta function.

From this formula and its derivative we can write the gap and its derivatives as
\be
\frac{\t \Delta(p_0)}{\t \Delta'(p_0)}= \frac{\lambda(p_0)}{f(p_0)}\;,\;\frac{\t \Delta''(p_0)}{\t \Delta'(p_0)}=\frac{1}{\lambda(p_0)}\left( \frac{f'(p_0)}{f(p_0)}\lambda(p_0)+f(p_0) - \lambda'(p_0) \right)\,.
\ee
Replacing these two expressions into the differential gap equation (\ref{eq:diff-gap-eq})
obtains
\be\label{eq:bcs-gap}
 \frac{d\lambda}{d \log p_0}= \frac{2\pi^2}{Z}\, \frac{du}{d \log p_0}- \frac{d \log Z}{d \log p_0}\,\lambda- \frac{1}{2\pi^2 N}\lambda^2\,.
\ee
This is the main result of the section, and we next show that it reproduces the correct 4-Fermi beta function.
(Let us note that this beta function has been derived before for the specific case of critical non-Fermi liquids in\cite{Raghu:2015sna}; and in fact that theory motivated our more general approach here. We will discuss this case in \S \ref{sec:NFL}.)

\begin{figure}[h!]
\centering
\includegraphics[width=1.4\textwidth]{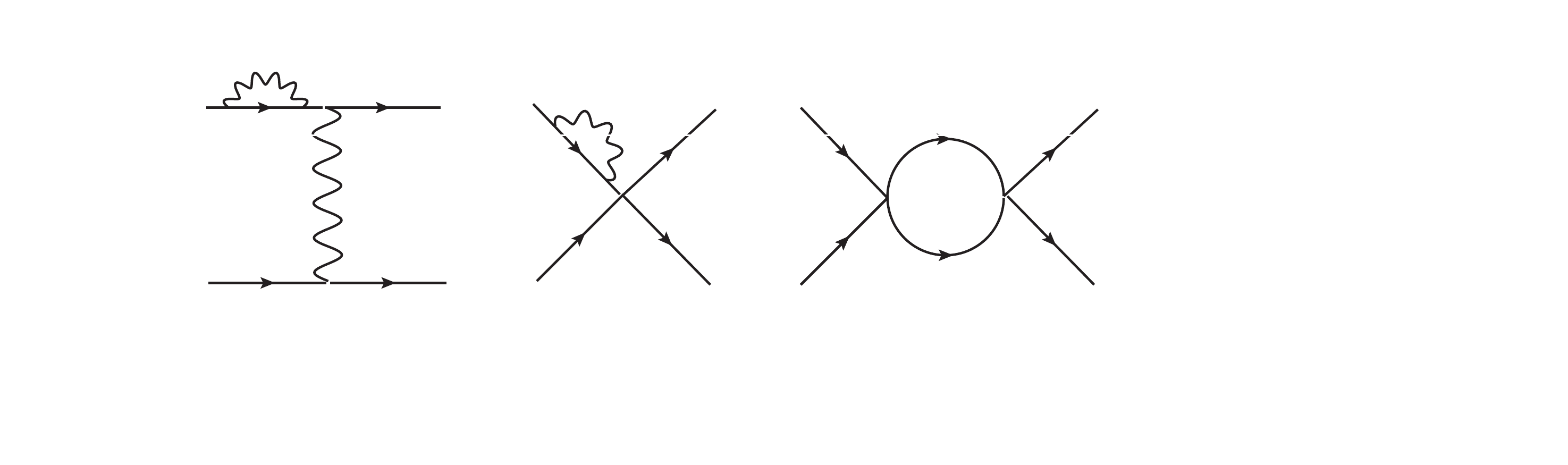}
\caption{One loop corrections that contribute to (\ref{eq:bcs-gap}): (dressed) tree-level boson exchange, anomalous dimension dressing, and BCS diagram. The straight line represents a fermion, while the wiggly line corresponds to a virtual boson.  }\label{fig:diagrams3}
\end{figure}

The first term is the tree level enhancement of the 4-Fermi coupling due to exchange of bosonic modes, where $u(p_0)$ is the boson propagator (\ref{eq:uD}) integrated over the Fermi surface. An example of this effect was first recognized by\cite{Son:1998uk} in the context of color superconductivity.
The second term is $2 \gamma \lambda$, where $\gamma$ is the fermion anomalous dimension, $2\gamma = - d \log Z/d \log p_0$. This term was first found for non-Fermi liquids in\cite{Fitzpatrick:2014efa, Raghu:2015sna}.  And the third term is the BCS fermion bubble that we already discussed. We show the corresponding diagrams in Fig. \ref{fig:diagrams3}. This gives all the right properties for the 4-Fermi coupling beta function in a finite density fermion system with general 4-Fermi kernel $u(p_0)$. 

We conclude that the RG beta function can be derived from the SDE equations by identifying the 4-Fermi coupling with
\be\label{eq:lambdagap}
\lambda(p_0) = \frac{2\pi^2 u'(p_0)}{Z(p_0)}\,\frac{\t \Delta(p_0)}{\t \Delta'(p_0)}\,.
\ee
This is valid for a generic $u(p_0)$, as long as the assumption of energy locality is satisfied. 
This ends our general analysis of fermions at finite density with arbitrary 4-Fermi kernel $u(p_0)$, obtained by integrating out bosonic modes. By assuming energy locality, we have derived the RG beta functions from the Eliashberg equation, obtaining the explicit map (\ref{eq:lambdagap}). This sheds new light on the physics of the frequency-dependent gap function, and on the behavior of the BCS coupling. In the rest of the work we will study these aspects in more detail in two important systems: phonon-mediated superconductivity, and non-Fermi liquids near a symmetry breaking transition at zero momentum. 

Finally, let us also note that one could use the exact RG (see\cite{Berges:2000ew} for a review), which keeps all the nonlocalities associated to mixing of high and low frequencies. A diagrammatic argument then shows that the gap function arises as the eigenvector of vanishing eigenvalue of the inverse fermionic 4-point function. An example of this for phonons was studied in\cite{2005PhRvB..72e4531T}.

\section{Phonon-mediated superconductivity}\label{sec:elph}

Having explained the equivalence of the RG and the linearized SDE equations under certain operating assumptions, we wish to put our knowledge to use.  In this section, to illustrate how the two approaches yield consistent solutions, we apply them to the classic problem of phonon-mediated superconductivity. In the process, we derive a new beta function for the 4-Fermi interaction that includes frequency dependent and strong coupling effects --as far as we know, this has not appeared before in the literature.

\subsection{Review of the electron-phonon system}\label{subsec:review}

In simple elemental metals such as Mercury and Niobium, superconductivity arises from electron-phonon coupling.  Eliashberg theory  arose as an attempt to explain the superconducting transition temperatures of elements such as Lead, which have stronger electron-phonon interactions.  See Refs. [\onlinecite{Eliashberg1960,Scalapino1966,McMillan1968,scalapino1969,Allen1975,Marsiglio2008}] for reviews (for an especially lucid treatment, see [\onlinecite{Scalapino1966}]).  As a consequence, there is a need to balance the effects of attraction from phonons with the enhanced scattering rate of normal-state quasiparticles that act to oppose superconductivity (referred to as a pair-breaking effect).  Instead of studying the detailed properties of actual phonons in metals, we study the simplest and ideal  case of optic phonons coupled to a Fermi liquid metal with a parabolic dispersion.  This will suffice to illustrate the equivalence between the RG and Eliashberg equations.  

We consider a metal in $d$ spatial dimensions with a spherical Fermi surface coupled to a single phonon mode,  
\begin{eqnarray}
S &=& \int \frac{d^d q dq_0}{(2 \pi)^{d+1}} \left[ \mathcal L_{\psi} + \mathcal L_{\phi} + \mathcal L_{\psi, \phi} \right] \nonumber \\
\mathcal L_{\psi} &=&  \psi^\dag_q \left( -i q_0 + v q_{\perp} \right) \psi_q - \frac{v_F}{2k_F^{d-1} A_{d-1}}\lambda \int \frac{d^d k}{(2 \pi)^d} \psi^\dag_{k+q}  \psi^\dag_{-k-q} \psi_{-q} \psi_{q} \nonumber \\
\mathcal L_{\phi} &=&\frac{1}{2}\, \phi_q\, \frac{ q_0^2 + \omega_q^2 }{\omega_q} \,\phi_{-q}\,,\;\; \omega_q = \omega_E \nonumber \\
\mathcal L_{\psi,\phi} &=& g \int \frac{d^d k}{(2 \pi)^d}  \psi^\dag_{k + q} \psi_{k} \phi_q\,.
\end{eqnarray}
We have adopted a spherical parametrization, writing a fermionic momentum as $\vec q =  \hat n(k_F+q_\perp)$ so that $q_\perp$ measures the radial distance towards the Fermi surface, and $\hat n$ is a unit vector. We will also denote tangential directions to the Fermi surface by $q_\parallel$.
The phonon is taken to be a simple dispersionless mode having a frequency $\omega_E$ (i.e. an Einstein phonon). 
The 4-Fermi coupling has been written in the BCS channel, and $\lambda>0$ corresponds to an attractive pairing. The factor $k_F^{d-1} A_{d-1}$ represents the area of the Fermi surface ($A_n$ is proportional to the area of the unit sphere in $n$ dimensions); it has been chosen such that quantum corrections to $\lambda$ are independent of the density of states.

The analysis will be performed in the limit where the masses of the nuclei are much larger than the electron mass. This implies that the sound speed is much smaller than the Fermi velocity, and that there is a hierarchy $\omega_E \ll E_F$. The Fermi scale $E_F$ will play the role of the microscopic cutoff. In this regime, Migdal's theorem implies that vertex corrections are suppressed and can be ignored. See\cite{abrikosov2012methods} for a review.

Before comparing the RG and SDE approaches to this problem, let us understand qualitatively the IR behavior: since $\omega_E$ is finite, we may simply integrate out the phonon, as well as all fermion modes with energy $E > \omega_E$,  to get a local BCS coupling for fermions close to the Fermi surface with energies $E \ll \omega_E$:
\begin{equation}
\lambda = \mu_* +c_d \frac{g^2 k_F^{d-1}}{v_F \omega_E}, \ \ \ \mu_* = \frac{\lambda_0}{1- \lambda_0 \log{\left[ E_F/\omega_E \right]}}\,.
\end{equation}
(The constant $c_d$ is determined below in (\ref{eq:lambdaE}).)
This will lead to a BCS instability if the net coupling is attractive and hence marginally relevant.  $\lambda_0$ is the  value of the repulsive BCS coupling (i.e. $\lambda_0<0$ given our conventions) at scales comparable to the ultraviolet cutoff --the Fermi energy $E_F$-- whereas $\mu^*$ is the renormalized BCS coupling at frequency $\omega_E$.  As we shall see below, the choice of $\mu^*$ will be equivalent to setting the value of the BCS coupling at the scale $\omega_E$. This estimate will be shown below to be accurate at weak coupling, while there will be large deviations from the BCS formula as the coupling becomes order one.

\subsection{SDE and RG frameworks}\label{subsec:RGphon}

We will now study the dynamics of the electron-phonon system using the SDE equations and the RG beta functions that follow from them according to \S \ref{sec:RG}. 

Let us recall first how to obtain the SDE equations for a Fermi surface coupled to a bosonic mode.
The derivation starts by allowing for a gap in the fermionic Lagrangian, written in the Nambu basis:
\be\label{eq:Lf}
L_f=- \psi_0(p)^\dag(iZ(p_0)p_0-v_0 p_\perp) \psi_0(p)+ Z(p_0) \Delta(p_0) \psi_0(p) \psi_0(-p)+\text{h.c.}
\ee
Neglecting vertex corrections, the standard fermionic Schwinger-Dyson equation reads
\be\label{eq:SD0}
G^{-1}(p)- [G^{(0)}]^{-1}(p) =-g^2 \int \frac{d^{d+1} q}{(2\pi)^{d+1}}\, D(q-p) G(q)\,,
\ee
where $D$ is the Landau-damped boson propagator and $G^{(0)}$ is the tree level fermion propagator (where $Z=Z_v=1$ and $\Delta=0$). The two components of (\ref{eq:SD0}) then reduce to
\bea\label{eq:SDE01}
\left(Z(p_0)-1\right)p_0&=& g_0^2 \int \frac{d^{d+1}q}{(2 \pi)^{d+1}} D(p-q)\,\frac{Z(q_0) q_0}{Z(q_0)^2(q_0^2+|\Delta(q_0)|^2)+v_0^2 q_\perp^2}\nonumber\\
Z(p_0)\Delta(p_0)&=&\frac{g_0^2}{N}\int \frac{d^{d+1}q}{(2 \pi)^{d+1}} D(p-q)\,\frac{Z(q_0) \Delta(q_0)}{Z(q_0)^2(q_0^2+|\Delta(q_0)|^2)+v_0^2 q_\perp^2}\,.
\eea

Since the phonon propagator 
 \begin{equation}
 D(p) = \frac{\omega_E}{p_0^2 + \omega_E^2}
 \end{equation}
 only depends on frequency, the SDE integrals will factorize trivially.
Doing the momentum integrals, one finds
\begin{eqnarray}
\left(Z(p_0) - 1\right) p_0 &=& \frac{1}{2 } \int \frac{ d q_0 }{(2 \pi)} u(p_0 - q_0) \frac{ q_0}{ \sqrt{ q_0^2 + \vert \Delta(q_0)\vert^2 } },  \nonumber \\
Z(p_0) \Delta(p_0) &=& \frac{1}{2} \int  \frac{d^{d+1} q}{(2 \pi)^{d+1}} u(p_0-q_0)\frac{ \Delta(q_0)}{ \sqrt{ q_0^2 + \vert \Delta(q_0)\vert^2 }}
\end{eqnarray}
 where 
 \begin{eqnarray}
 u(q_0 - p_0) &=& \frac{g_0^2}{v_0}\int \frac{d^{d-1} q_{\parallel} }{(2 \pi)^{d-1}}D(p-q) = \frac{g_0^2}{v_0}\frac{k_F^{d-1}}{(2 \pi)^{d-1}} \frac{2 \pi^{(d-1)/2}}{\Gamma((d-1)/2)} \left[ \frac{\omega_E}{(q_0 - p_0)^2 + \omega_E^2} \right] \nonumber \\
 &\equiv& \alpha_0 \frac{\omega_E}{(q_0 - p_0)^2 + \omega_E^2}
 \end{eqnarray}
These are of the form (\ref{eq:El1}). For an estimate of the pairing scale, we work near the normal state and study the linearized equations (\ref{eq:El2})

\subsubsection{Fermion self-energy}

The wavefunction renormalization factor can be obtained from the first expression above; since the pairing scale is much smaller than the UV cutoff, it can be approximated as zero at a first pass:
\begin{eqnarray}\label{eq:Zphonon}
Z(p_0) &\approx& 1 + \frac{\alpha_0 \omega_E}{2p_0} \int_{-\Lambda_0}^{\Lambda_0} d q_0 \frac{{\rm sgn}(q_0) }{(p_0 - q_0)^2 + \omega_E^2}  \nonumber \\
&=& 1 + \frac{\alpha_0}{p_0} \arctan{\left[ \frac{p_0}{\omega_E} \right]}\,.
\end{eqnarray}
This result encodes different physics, depending on whether the frequency is above or below the phonon scale.

Consider first the high frequency regime, $p_0 \gg \omega_E$. The fermion self-energy becomes
\be
\Sigma(p_0) = (Z(p_0)-1) p_0 \approx \frac{\pi}{2}\alpha_0\,.
\ee
This is the expected contribution from a homogeneous density of massless impurities, consistent with the fact that the phonon mass can be neglected at high frequencies. For $p_0 < \omega_E$, on the other hand, the phonons become massive and decouple, leaving behind a constant correction to $Z(p_0)$, namely $\Sigma(p_0) \approx \frac{\alpha_0}{\omega_E} p_0$.

The full result (\ref{eq:Zphonon}) also describes the intermediate regime $p_0 \sim \omega_E$, where there is no simple scaling form for $\Sigma$. This reflects the decoupling of massive modes in a physical RG scheme; it will give rise below to interesting deviations from the BCS superconductor.

\subsubsection{Superconductivity and 4-Fermi coupling}

Next consider the equation for the gap $\Delta$.    As before, we employ the notion of locality in (\ref{eq:local1}), so that the gap equation becomes equivalent to the differential equation
(\ref{eq:diff-gap-eq}). Following the same steps in the previous section, we arrive at the RG equation equivalent to the linearized SDE equations:
 \begin{equation}
 \label{RGeq_phonon}
 \frac{ d \lambda}{d \log{p_0} } = \frac{2 \pi^2}{Z} \frac{d u}{d \log{p_0} } - \frac{ d \log{Z}}{d \log{p_0}} \lambda - \frac{1}{2 \pi^2} \lambda^2\,.
 \end{equation}
We apply this to the following functions:
 \begin{equation}\label{eq:phonon-uZ}
 u(p_0) = \alpha_0 \frac{\omega_E}{(p_0)^2 + \omega_E^2},  \ \ \ Z(p_0)  = 1 + \frac{\alpha_0}{p_0} \arctan{\left[ \frac{p_0}{\omega_E} \right]}\,.
 \end{equation}
 
 
We solve the equation above numerically by working in logarithmic coordinates.  Defining $ |p_0| = \omega_E e^{-x}$, 
 \begin{equation}
u(x) = \frac{\alpha_0}{\omega_E} \frac{1}{1 + e^{-2x}},  \ \ Z(x) = 1 + \frac{\alpha_0}{\omega_E} e^x \arctan{\left( e^{-x} \right)}
 \end{equation}
 the RG equation is
 \begin{equation}
 \frac{ d \lambda}{d x } = \frac{2 \pi^2}{Z} \frac{d u}{d x} - \frac{ d \log{Z}}{d x} \lambda + \frac{1}{2 \pi^2} \lambda^2\,.
 \end{equation}
For simplicity, we solve this equation with the initial condition $\lambda(x=0) = 0$, which is equivalent to setting $\mu^* = 0$ in (\ref{eq:BCSlambda}). 

\begin{figure}[h!]
\centering
\includegraphics[width=0.49\textwidth]{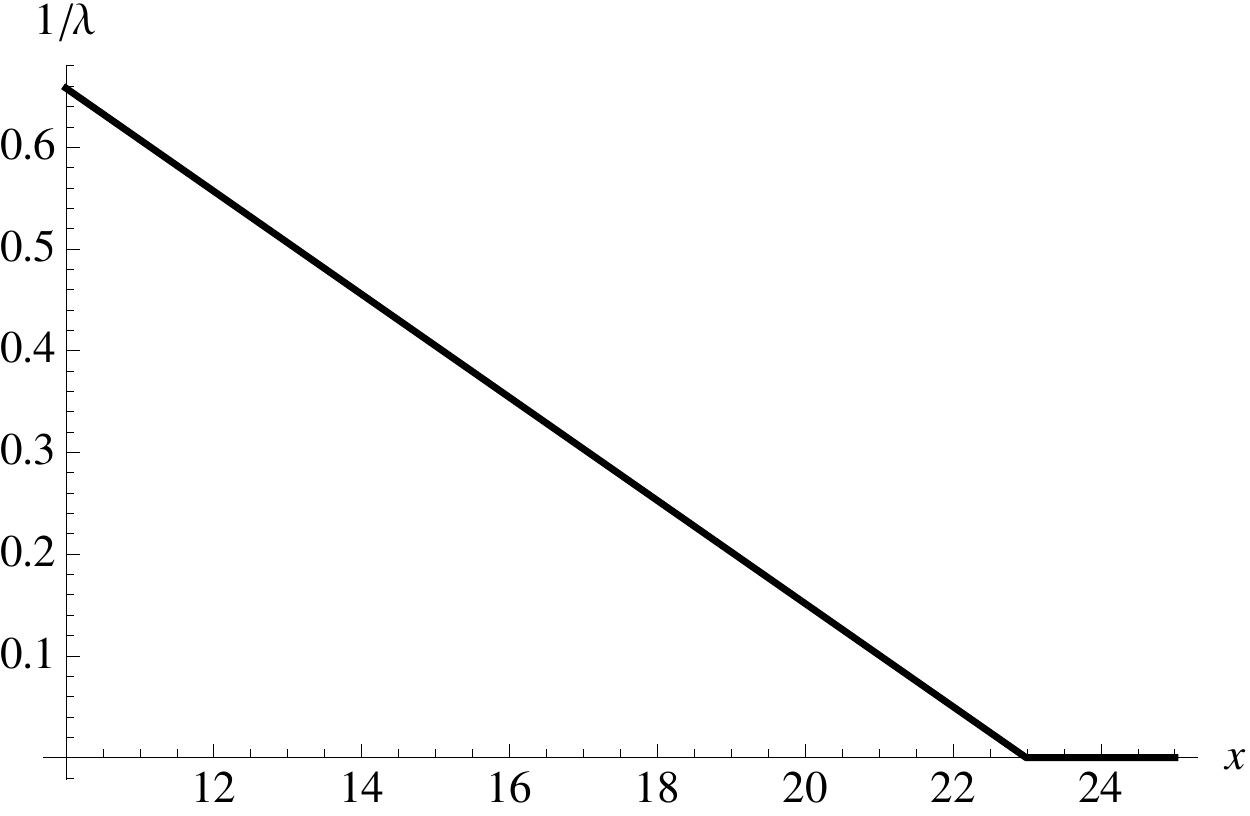}
\includegraphics[width=0.49\textwidth]{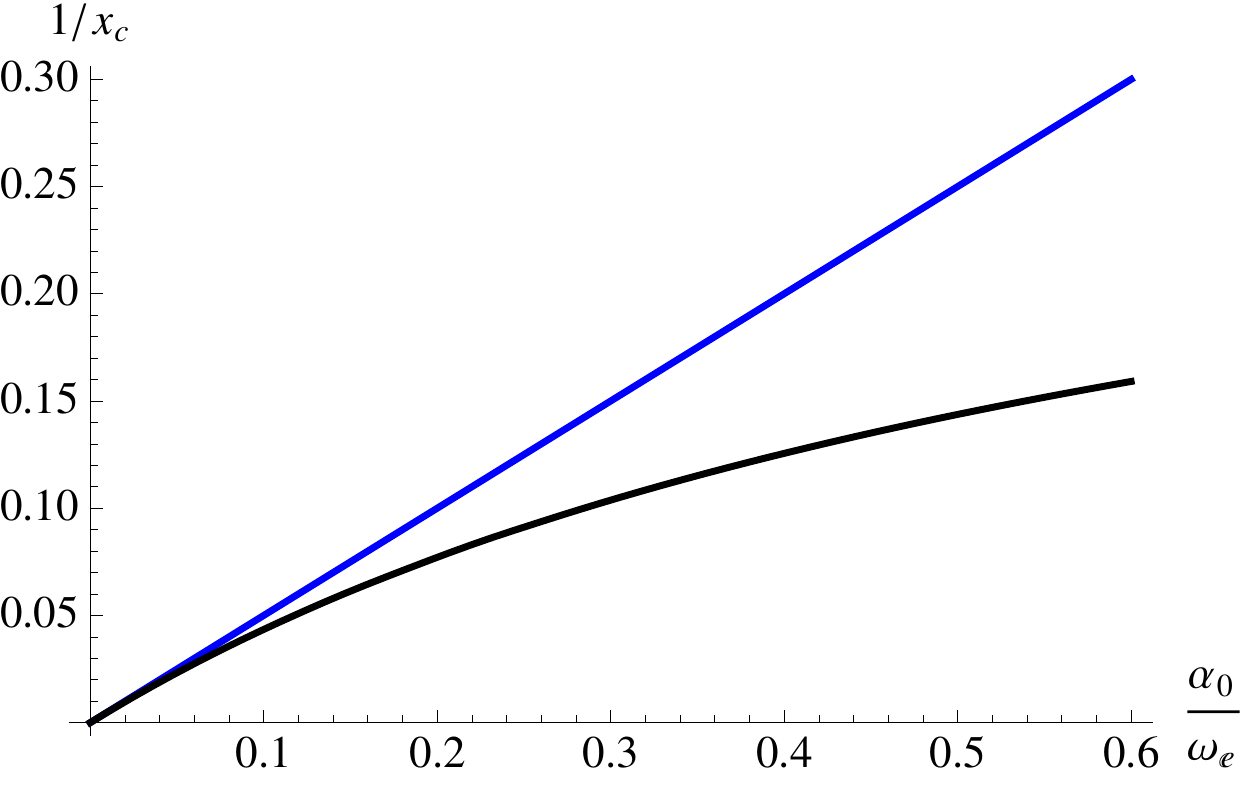}
\caption{Left: Plot of $1/\lambda(x)$ as a function of $x$ for $\alpha_0/\omega_E=0.1$ showing a zero (equivalent to a divergence of $\lambda$ at a critical value $x_c$). Right: The critical value $x_c$ is obtained for various values of $\alpha_0/\omega_E$ (black curve).  A linear relationship exists between $1/x_c$ and $\alpha_0/\omega_E$ when the latter ratio is small, consistent with the BCS prediction (blue line). At larger couplings, quantum effects introduce deviations from the BCS behavior.}\label{fig:phonon1}
\end{figure}

The left panel in Fig. \ref{fig:phonon1} shows the behavior of the numerical solution $\lambda(x)$, for $\alpha_0/\omega_E=0.1$.  We have plotted the inverse of $\lambda(x)$, which enables us to find the critical value $x_c$ at which $\lambda(x_c)$ diverges.  We identify this as the instability scale $\Delta_*$.  We repeat this for various values of $\alpha_0/\omega_E$, showing the dependence of $x^{-1}_c(\alpha_0/\omega_E)$ in the right panel of Fig. \ref{fig:phonon1} (black curve).  

When the ratio $\alpha_0/ \omega_E \ll 1$, there is a linear relationship between these two quantities.  This is entirely consistent with the predictions of BCS,
\begin{equation}
\frac{\Delta_*}{\omega_E} = e^{-x_c} \simeq  e^{-2\pi^2/\lambda(\omega_E)}\,.
\end{equation}
For this, notice that integrating out the phonon gives rise to
\be\label{eq:lambdaE}
\lambda(\omega_E) =\frac{2\pi^2}{Z(\omega_E)} u(\omega_E) \approx \pi^2 \frac{\alpha_0}{\omega_E}
\ee
where the approximate equality holds for small $\alpha_0/\omega_E$. This can also be seen directly in the beta function: it is obtained by integrating the tree level term between the UV cutoff and $\omega_E$. Hence $x_c^{-1} \approx \frac{1}{2} \frac{\alpha_0}{\omega_E}$ in the BCS regime $\alpha_0/\omega_E \ll 1$, in agreement with the numerical result.

The exponentially small dependence of the gap scale on the effective interaction is consistent with the marginally relevant BCS coupling of a Landau Fermi liquid. However, only the third term of the more general RG equation in Eq. (\ref{RGeq_phonon}) for $\lambda$ reflects the marginal relevance.  It must follow that the other two terms must be parametrically small in $p_0/\omega_E$, i.e. at low energies near the Fermi liquid fixed point.  And indeed, an explicit calculation shows that these terms are of order $\frac{\alpha_0}{\omega_E} (\frac{p_0}{\omega_E})^2$. Thus, at low energies and weak coupling, the BCS approximation holds self-consistently. 

On the other hand, as $\alpha_0/\omega_E$ increases, the behavior of the gap deviates from the BCS prediction. In particular, the curve including larger values of $\alpha_0/\omega_E$, is well approximated by a function of the form $x_c^{-1} =c_0 \frac{\alpha_0/\omega_E}{c_1 + \alpha_0/\omega_E}$ with constants $c_0$ and $c_1$. Thus, as the coupling increases, the gap changes from a BSC-like dependence $\Delta_* \sim e^{-1/\lambda_0}\omega_E$ to a strong-coupling behavior $\Delta_* \sim e^{-(const+\lambda_0)/\lambda_0}\omega_E$. This is due to the first two terms in the beta function (\ref{RGeq_phonon}), which encode effects from tree-level exchange and anomalous dimension dressing. This type of behavior is familiar from studies of phonon-mediated superconductivity at strong coupling, such as the McMillan formula\cite{McMillan1968}.  It would be interesting to understand these effects in more detail within a controlled strong-coupling expansion --something that is beyond the scope of the present work, but which we hope to address in the future.

We have shown that the RG equation obtained from the SDE analysis for the phonon problem yields sensible estimates for the pairing scale.  However, we used the idea of locality of $u(p_0 - q_0)$ without justifying this approximation in this section.  In the next section, we show that the solutions using the local approximations deviate very little from the exact solutions of the SDE equations for the phonon problem, which is the {\it a posteriori} justification for assuming locality of the SDE kernel.

\subsection{Numerical analysis of the local approximation}

The integral SDE equation in the linearized regime takes the form (\ref{eq:El2}),
where the self-energy and the phonon-kernel are given by (\ref{eq:phonon-uZ}).
In this section we numerically check the claim that the solutions can be approximated by those of the corresponding differential equation (\ref{eq:local2}); then the map to the RG description will be valid.

It is convenient to go to the log variable by defining $|p_0|=\omega_E\;e^{-x}$, and re-write the differential and integral equations as 
\bea\label{eq: phonon_gap_x}
&& \t\Delta''(x)+2\tanh{(x)}\t\Delta'(x)+\frac{g_1}{2}\frac{\t\Delta(x)}{\cosh^2{(x)}\left(1+g_1e^x\tan^{-1}\left(e^{-x}\right)\right)}=0\,,\\
&&\t\Delta(x)=\frac{g_1}{2}\int^{x_\text{gap}}_{x_\Lambda}dx'\left\lbrace \frac{1}{\left(e^{-x}-e^{-x'}\right)^2+1}+\frac{1}{\left(e^{-x}+e^{-x'}\right)^2+1}\right\rbrace\frac{\t\Delta(x')}{1+g_1 e^{x'}\tan^{-1}\left(e^{-x'}\right)}\,,\nonumber
\eea
where $g_1=\frac{\alpha_0}{\omega_E}$ is the only dimensionless parameter that the system depends on. We numerically obtain the differential equation solution $\t\Delta_\text{diff}(x)$, under the UV boundary condition $\t\Delta_\text{diff}(x_\Lambda)=0$, from which we can solve the gap in the  $x$-variable $\t\Delta_\text{diff}'(x_\text{gap})=0$. We can then check whether $\left\lbrace \t\Delta_\text{diff}(x), x_\text{gap}\right\rbrace$ satisfies the integral equation, schematically written as: 
\be
\t\Delta_\text{diff}(x) = F\circ \t\Delta_\text{diff}(x) \,.
\ee

It does not seem possible to have closed-form solutions to (\ref{eq: phonon_gap_x}). But numerically we only need the correct initial conditions at the UV. To do this, we obtain approximate solutions near $x\to -\infty$, where the equation simplifies: 
\be
\t\Delta''(x)-2\t\Delta'(x)+2g_1 e^{2x}\t\Delta(x)=0 \,.
\ee
The general solution is given in terms of Bessel functions,
\be 
\t\Delta(x)=C_1\,e^x \,J_1(\sqrt{2g_1}e^x) +C_2\,e^x\, Y_1(\sqrt{2g_1}e^x)\,.
\ee 
Taking the UV boundary condition $\t\Delta(-\infty)=0$ selects $C_1=0$. This provides the UV boundary condition to numerically solve the differential equation. 

\begin{figure}[h!]
\centering
\includegraphics[width=0.6\textwidth]{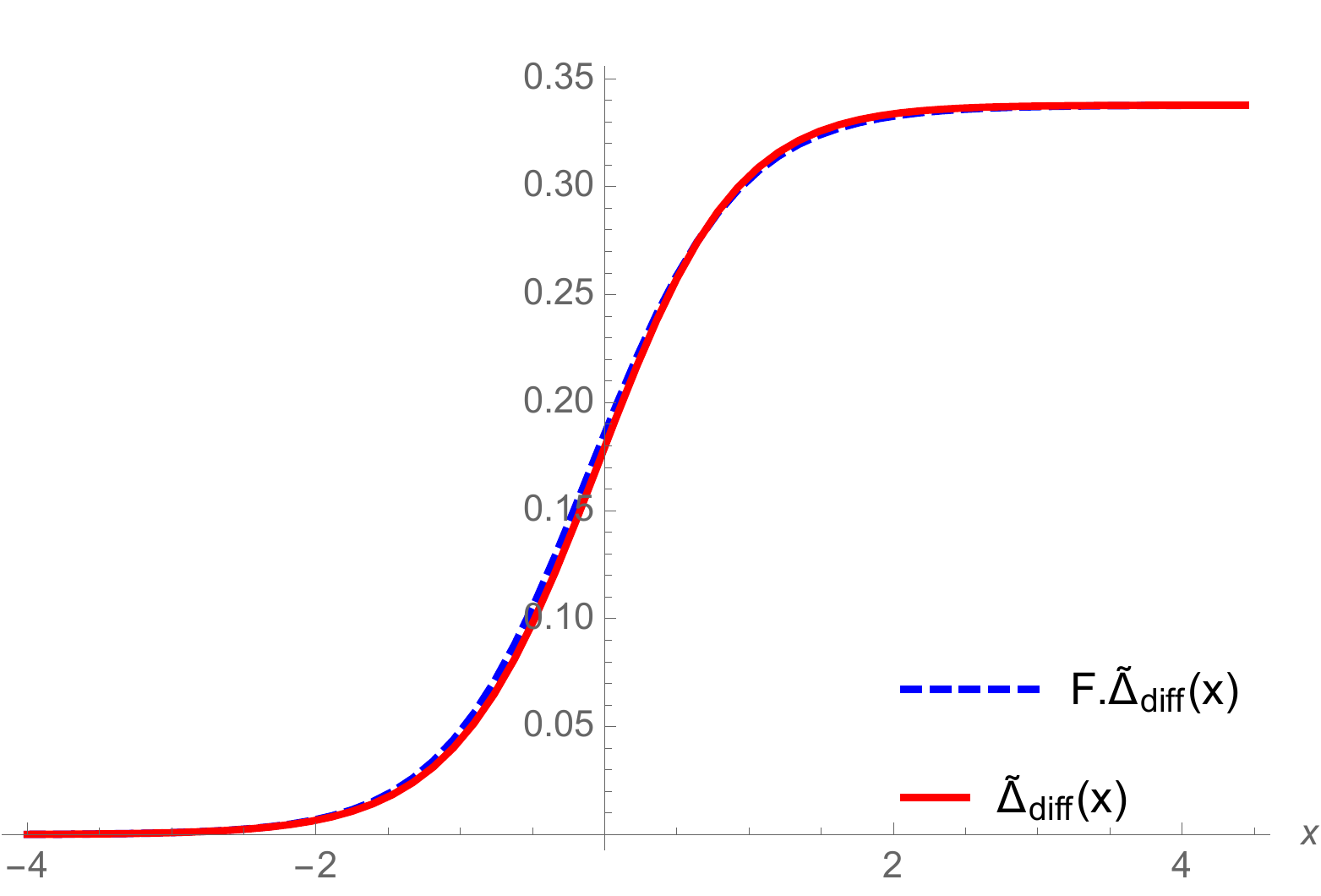}
\caption{$\t\Delta_\text{diff}(x)$ against $F\circ \t\Delta_\text{diff}(x)$, for which we picked $x_\Lambda=-12$, and $g_1=0.2$. The gap is solved to be at $x_\text{gap}=1.7006$.}\label{fig: phonon_check}
\end{figure}

In Fig. \ref{fig: phonon_check} we plot the solution to the differential equation $\t\Delta_\text{diff}(x)$ against $F\circ \t\Delta_\text{diff}(x)$; coincidence of the two indicates a solution to the integral SDE equation. We have found, for generic values of the parameters, that the two coincide very well. We conclude that the local approximation is well-justified in the phonon-mediated system. 

\section{Non-Fermi liquids with critical bosons}\label{sec:NFL}

In this section we study non-Fermi liquids coupled with critical bosons, which arise in the proximity of symmetry breaking transitions with zero momentum.
 We first review previous results using the RG methods of\cite{Fitzpatrick:2014efa, Raghu:2015sna}, and then derive the Eliashberg-Schwinger-Dyson equations for these systems. These fall into the general framework of \S \ref{sec:RG}, and hence we demonstrate the equivalence of the RG and SDE results for non-Fermi liquids. In \S \ref{sec:interplay} we will apply these results to the interplay between superconductivity and quantum criticality.

\subsection{Review of the model and RG results}\label{subsec:nflreview}

Let us briefly introduce our model of large $N$ non-Fermi liquids and their RG description. More details can be found in\cite{Raghu:2015sna}. We consider theories with an $SU(N)$ global symmetry, with the fermions $\psi_i$ transforming in the fundamental, coupled to a critical $N\times N$ scalar $\phi_i^j$ in the adjoint representation\footnote{This limit  is different from the usual large N limits that ons commonly sees in the literature on metallic criticality.   Usually, one studies the case of a large number of fermionic flavors and only one bosonic degree of freedom\cite{Polchinski1994, Altshuler1994, Lee2009, Metlitski}}. We also work near three spatial dimensions (the critical dimension), performing an $\epsilon$ expansion, $d=3-\epsilon$. 

The euclidean lagrangian takes the following form:
\be\label{eq:L1}
L=\frac{1}{2}\tr \left((\partial_\tau \phi)^2+ (\vec 
\nabla \phi)^2 \right)+ \psi_i^\dag \left(\partial_\tau+ \varepsilon(i\vec \nabla)-\mu_F
\right)\psi^i +L_\text{Yuk}+L_\text{BCS}\,,
\ee
with boson-fermion interaction
\be\label{eq:LY}
L_\text{Yuk}=\frac{g}{\sqrt{N}} \phi^i_j \psi^\dag_i \psi^j
\ee
and 4-fermion interaction along the BCS channel
\be\label{eq:LBCS}
L_\text{BCS}=-\frac{1}{2k_F^2} \frac{v \lambda}{N}\, \psi^\dag_i(p+q)\psi^j(p)\psi^\dag_j(-p-q)\psi^i(-p)\,.
\ee
We work below the Landau damping scale $M_D^2 \approx \frac{ k_F^{d-1}}{2\pi v} \frac{g^2}{ N}$, where the matrix scalar $\phi_i^j$ has an emergent $z=3$ scaling. The boson propagator in the regime where Landau damping dominates is given by
\be
D(q_0, q) = \frac{1}{q^2+M_D^2 \frac{|q_0|}{v |q|}}\,.
\ee

In\cite{Raghu:2015sna} we have constructed a scaling theory for this problem including the running BCS coupling.  There, we found the following RG equations
\bea\label{eq:betafcs1}
&&2\gamma= -\mu\frac{d\log Z}{d\mu}=\alpha\nonumber\\
&&\mu\frac{d\alpha}{d\mu} =-\frac{\epsilon}{3} \alpha + 2\gamma \alpha\,, \nonumber
\eea
where $\mu$ is the RG energy scale, and $\alpha= g^2/(12 \pi^2)$.
This system admits a non-Fermi liquid fixed point~\cite{Torroba:2014gqa, Raghu:2015sna},
\be\label{eq:qcp}
\alpha_*=\frac{\epsilon}{3}\;,\;\gamma_*=\frac{\epsilon}{6}
\ee
that is perturbatively controlled at small $\epsilon$ and large $N$. An important property of the model is that there is no restriction on $\epsilon N$; this will not be the case of the vector large N limit briefly discussed in  \S \ref{subsec:BKT}. Furthermore, the BCS beta function was found to be
\be\label{eq:betaBCS}
\mu\frac{d\l}{d\mu} = -2 \pi^2 \alpha+2\gamma \l - \frac{\l^2}{2\pi^2 N}\,.
\ee
We recognize the similarity with (\ref{eq:bcs-gap}), and we will demonstrate shortly that both are in fact the same.

The quantum contributions of the NFL fixed point appear in the tree level and anomalous dimension terms of (\ref{eq:betaBCS}). For $\epsilon N \ll 1$, NFL effects are small, and the dynamics is very similar to the color superconductor of\cite{Son:1998uk}. However, as $\epsilon N$ is increased, the quasiparticles become more incoherent, pair-breaking effects increase, and superconductivity tends to disappear. Eventually, superconductivity is completely extinguished through a BKT-like transition, and for large $\epsilon N$ the system reaches a new quantum critical regime with constant BCS coupling. Our goal here is to capture these effects from the SDE approach.

\subsection{SDE equations}\label{subsec:eqs}

Let us study the non-Fermi liquids (\ref{eq:L1}) in terms of the SDE approach. It will be shown that the resulting gap equation is in agreement with the RG approach. 

The SDE equations are more transparently derived in terms of bare couplings. We will distinguish these from the running RG couplings by using a subindex `$0$'. For simplicity, the bare 4-Fermi coupling is chosen to vanish; the bare Yukawa coupling and velocity are denoted by $g_0$ and $v_0$.

The derivation proceeds as in \S \ref{subsec:RGphon}. The fermionic Lagrangian is (\ref{eq:Lf}), with additional
contractions of flavor $SU(N)$ indices. For our purpose, it will be sufficient to consider a very symmetric symmetry breaking pattern, where $\Delta_{ij}$ is proportional to the symplectic matrix $i \sigma_2 \otimes \mathbf 1_{N/2}$, and $N$ is even. At large $N$, the two components of (\ref{eq:SD0}) then reduce to
\bea\label{eq:SDE1}
\left(Z(p_0)-1\right)p_0&=& g_0^2 \int \frac{d^{d+1}q}{(2 \pi)^{d+1}} D(p-q)\,\frac{Z(q_0) q_0}{Z(q_0)^2(q_0^2+|\Delta(q_0)|^2)+v_0^2 q_\perp^2}\nonumber\\
Z(p_0)\Delta(p_0)&=&\frac{g_0^2}{N}\int \frac{d^{d+1}q}{(2 \pi)^{d+1}} D(p-q)\,\frac{Z(q_0) \Delta(q_0)}{Z(q_0)^2(q_0^2+|\Delta(q_0)|^2)+v_0^2 q_\perp^2}\,.
\eea

A crucial point is that the diagrams contributing to $Z(p)$ are planar, while the contributions to the gap are nonplanar in the $SU(N)$ theory, and hence suppressed by $1/N$. This will enhance non-Fermi liquid effects on the low energy dynamics.

Finally, let us note an important property of (\ref{eq:SDE1}) which will simplify the following analysis. Due to the $z=3$ scaling of the boson momentum, the dependence of $D(p-q)$ on $q_\perp$ can be neglected. The integrals over $q_\perp$ and $q_\parallel$ then factorize between the boson and fermion lines. Quantum corrections are then controlled by the integral of the boson propagator over the tangential Fermi surface directions,
\be\label{eq:u1}
u(p_0-q_0) \equiv 6\pi \alpha_0  \int \frac{d^{2-\epsilon}q_\parallel}{(2\pi)^{2-\epsilon}}\,D(q-p)=\frac{3\alpha_0}{\epsilon}\frac{1}{\left(M_D^2|p_0-q_0|\right)^{\epsilon/3}} +\mathcal O(\epsilon^0)\,.
\ee
Similar power-law kernels arise in various other models, and our conclusions will also hold for them as long as $\epsilon$ remains small. The linearized Eliashberg equations are again given by (\ref{eq:El2}).
One can immediately compute the wavefunction factor:
\be
Z(p_0)\approx 1+\frac{3\alpha_0}{\epsilon}\,\frac{1}{(M_D^2 |p_0|)^{\epsilon/3}}\,,
\ee
valid at small $\epsilon$ and for $|p_0|> \Delta_*$.

Therefore, the NFL model with the kernel (\ref{eq:u1}) falls into the class of systems studied in \S \ref{sec:RG}, and we can directly apply those results. Implementing the local approximation, and changing variables to
\be\label{eq:x-variable}
p_0=\left(\frac{3\alpha_0}{\epsilon}\right)^{3/\epsilon} M_D^{-2 }e^{-3 x/\epsilon}\,,\,g_1 =\frac{3}{\epsilon N}\,,
\ee
obtains the differential equation for the gap
\be\label{eq:gapZeq}
(1+e^x)(\t \Delta''(x)-\t \Delta'(x))+g_1 e^x \t \Delta(x)=0\,.
\ee
In terms of these variables, the physical gap corresponds to $x_*$ defined by
\be
\Delta_*=\left(\frac{3\alpha_0}{\epsilon}\right)^{3/\epsilon}M_D^{-2} e^{-3 x_*/\epsilon}\,.
\ee
This makes manifest that the only dependence on $3\alpha_0/\epsilon$ is in the overall factor, while $x_*$ depends only on $g_1$. The UV boundary condition (\ref{eq:UVbc}) reads
\be
\t \Delta'(x_\Lambda)-\t \Delta(x_\Lambda)= 0\,,
\ee
($x_\Lambda<0$ corresponds to the frequency cutoff $\Lambda_0$ above)
while the gap is determined by
\be
 \t \Delta'(x_*)=0\,.
\ee

We can now apply the proposed relation between variables in \S \ref{subsec:RGderivation}:
\be\label{eq:RG-SD1}
\lambda(x)=2\pi^2 \alpha_1(x) \,\frac{\t \Delta(x)}{\t \Delta'(x)}\,.
\ee
One can verify that the BCS beta function and Eliashberg equations of this system are indeed equivalent under this identification:
\bea\label{eq:betalx}
&&\lambda'(x)=-2\pi^2 \alpha_1(x)+\alpha_1(x)\lambda(x)-\frac{g_1}{2\pi^2}\lambda(x)^2 \nonumber\\
&&\t \Delta''(x)=\t \Delta'(x)-g_1 \alpha_1(x)\t \Delta(x)\,,
\eea
Note that at the physical gap, $\t \Delta'(x_*)=0$, leading to the instability $\lambda \to \infty$. In this way, we have established the equivalence between the Eliashberg and RG formulas for NFLs. It remains to verify the validity of the local approximation, to which we turn next.


\subsection{Numerical analysis}\label{subsec:numerics}

The last step to establish the equivalence between the gap and RG approaches is to show that the approximations leading to the differential equation (\ref{eq:gapZeq}) are valid. The underlying physical assumption that needs to be tested is whether the energy-scale locality of the RG is preserved by the integral gap equation. We will do this numerically, as we have not been able to solve the integral equation analytically. Ref. [\onlinecite{2016arXiv160601252W}] found evidence for nonlocalities, but our results show no sign of them, and an excellent agreement between the RG and Eliashberg results.

More specifically, we demonstrate that the solutions to (\ref{eq:gapZeq}) satisfy the linearized Eliashberg equation (\ref{eq:El2}), which in terms of the $x$ variable can be written as
\bea\label{eq:SDlinearX}
\tilde{\Delta}(x)&=&\frac{g_1}{2}\int^{x_*}_{x_\Lambda}dx' u(x,x')\frac{\tilde{\Delta}(x')}{1+e^{x'}}\nonumber\\
 u(x,x')&=& | e^{-\frac{3x} {\epsilon}}+e^{-\frac{3x'} {\epsilon}}|^{-\frac{\epsilon}{3}} +  | e^{-\frac{3x} {\epsilon}}-e^{-\frac{3x'} {\epsilon}}|^{-\frac{\epsilon}{3}}\,.
\eea
To make the numerics comparison concrete, we specify a finite UV cutoff $x_\Lambda<0$. It enters the Eliashberg equation as a lower-cutoff for the $x$ integral; for the differential equation it imposes the boundary condition
\be\label{eq:UVbdry}
\t\Delta'(x_\Lambda)-\t\Delta(x_\Lambda)=0 \,.
\ee

There are two linearly independent hypergeometric solutions to (\ref{eq:gapZeq}), and a general solution can be written as a linear combination:
\bea\label{eq:Diff_sol}
&&\t\Delta_\text{dif}(x)= e^x \,_2F_1\left(\frac{1}{2}-\frac{1}{2}\sqrt{1-4g_1},\frac{1}{2}+\frac{1}{2}\sqrt{1-4g_1},2,-e^x\right)\\
&+& C_\Lambda\text{MeijerG}\left(\left\lbrace \lbrace\rbrace,\left\lbrace\frac{3}{2}-\frac{1}{2}\sqrt{1-4g_1},\frac{3}{2}+\frac{1}{2}\sqrt{1-4g_1}\right\rbrace\right\rbrace,\left\lbrace \lbrace 0,1\rbrace,\lbrace\rbrace\right\rbrace, -e^x\right)\,.\nonumber
\eea 
Recall that the overall scale is not fixed in the linearized approximation.
For a given UV cutoff $x_\Lambda$ and $g_1$, we first compute the numerical coefficient $C_\Lambda$ by imposing the UV boundary condition (\ref{eq:UVbdry}). This turns out to be exponentially small in $|x_\Lambda|$, with the hypergeometric term already satisfying (\ref{eq:UVbdry}) to a very good approximation. The physical gap $x_*$ is then obtained by solving $\t\Delta'_{\text{dif}}(x_*)=0$. For our purpose here, we found it more convenient to solve the differential system numerically; (\ref{eq:Diff_sol}) will be useful to understand NFL effects below.

\begin{figure}[h!]
\centering
\includegraphics[width=0.6\textwidth]{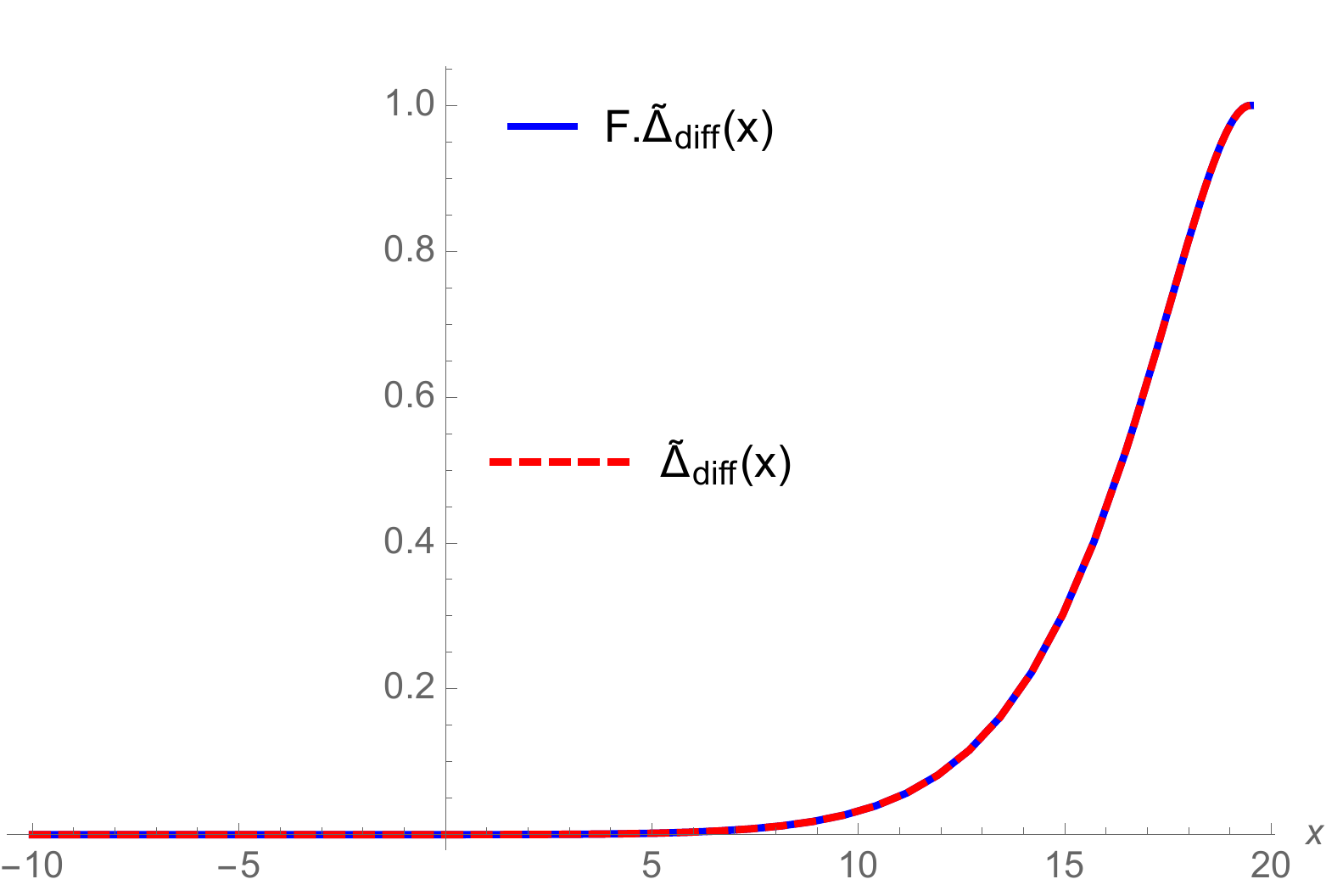}
\caption{Comparison between $\t\Delta_\text{dif}(x)$ (red, dashed) and $F\circ \t\Delta_\text{dif}(x)$ (blue), normalized by their values at $x_*$, with $x_\Lambda=-10,\;g_1=0.27,\;\epsilon=0.1$, for which $x_*\approx 19.49$. In these variables, the NFL scale is at $x=0$, and the UV cutoff at $x_\Lambda=-10$.}\label{fig:diff-integral}
\end{figure}

We check if such a solution solves the Eliashberg equation by comparing $\t\Delta_\text{dif}(x)$ and 
\be
F\circ \t\Delta_\text{dif}(x)=\frac{g_1}{2}\int^{x_*}_{x_\Lambda}dx'u(x,x')\frac{\t\Delta_\text{dif}(x')}{1+e^{x'}}\,.
\ee 
Equality between the two in the range $x_\Lambda\leq x\leq x_*$ indicates that the Eliashberg equation is fully satisfied. In Fig. \ref{fig:diff-integral} we show one example of such numerical comparison, and conclude that the differential equation captures the full solution to the integral Eliashberg equation to very good precision. This completes our analysis of the relation between the RG and Eliashberg approaches.

\section{Interplay between superconductivity and quantum criticality}\label{sec:interplay}

In this last section we consider some consequences and applications of our results. We explore NFL effects on superconductivity; there is a BKT-like transition through which superconductivity is destroyed, and the system transitions to a quantum critical state with finite BCS interaction. We also discuss some applications of this to the normal state of unconventional metals. 

\subsection{NFL effects on superconductivity}\label{subsec:BKT}

Let us explore NFL effects on the superconducting parameter using the results from the Eliashberg equations.
From the differential equation approach, the general solution takes the form (\ref{eq:Diff_sol}). In the limit $x_\Lambda\to-\infty$, $C_\Lambda\to 0$, we have that 
\be\t\Delta(x)\approx e^x \,_2F_1\left(\frac{1}{2}-\frac{1}{2}\sqrt{1-4g_1},\frac{1}{2}+\frac{1}{2}\sqrt{1-4g_1},2,-e^x\right)\,.
\ee 
Using a few identities among hypergeometric functions, we deduce the condition for the physical gap
\be\label{eq:physica_gap}
P_{-\frac{1}{2}+i\sqrt{g_1-1/4}}(1+2e^{x_*})=0 \,,
\ee
where $P_{l}$ is the associated Legendre polynomial. Clearly, pair-breaking contributions become strong as $g_1 \to 1/4$.

We now argue that $\t \Delta$ vanishes via an infinite order transition at $g_1=1/4$. For this, expand (\ref{eq:physica_gap}) in small $\delta_g =\sqrt{g_1-1/4}\ll 1$, with $e^{x_*}\gg 1$:
\be
\Gamma\left(\frac{1}{2}+i\delta_g\right)\Gamma(-i\delta_g)+16^{i\delta_g}\Gamma\left(\frac{1}{2}-i\delta_g\right)\Gamma(i\delta_g)e^{2i\delta_g x_*}=0 \,.
\ee
One can then solve for $x_*$:
\be\label{eq:BKT-SD}
x_*(\delta_g)=C(\delta_g)+\frac{n\pi}{\delta_g},\;\lim_{\delta_g\to 0}C(\delta_g)=-\ln{16}\approx-2.77 \,,
\ee
where $n$ is a positive integer. The physical gap corresponds to the first such solution encountered coming from $x_\Lambda<0$, namely $n=1$. Relating $x_*$ back to the physical frequency, we conclude that near the transition $g_1=1/4$, the gap vanishes like: 
\be\label{eq:xc}
\lim_{\Lambda\to\infty} \Delta_*(g_1) =  \left(\frac{3\alpha_0}{\epsilon}\right)^{3/\epsilon}\Lambda^{-2}e^{-3x_*(g_1)/\epsilon},\;x_*(g_1)=-2.77+\frac{\pi}{\sqrt{g_1-1/4}}\,.
\ee
Therefore, as $g_1 \to 1/4$, the SC gap vanishes through a BKT-type phase transition. 
For completeness, we also confirmed numerically the behavior of (\ref{eq:BKT-SD}) for finite UV cutoff $x_\Lambda$, finding very good agreement with the result corresponding to infinite UV cutoff.

Let us summarize the behavior we have found. For sufficiently small $\epsilon N$ (namely $g_1=\frac{3}{\epsilon N}>\frac{1}{4}$), the superconducting phase is stable. The gap (\ref{eq:xc}) agrees with Son's result\cite{Son:1998uk}. As $g_1 \to 1/4$, $\Delta_*$ and all its derivatives vanish --this is a BKT type transition through which superconductivity is destroyed. For smaller $g_1$, i.e. $\epsilon N > 12$, there is no superconducting instability. This is consistent with the results from the RG analysis in \cite{Raghu:2015sna}.

It is also interesting to ask whether it is possible to have large NFL effects in a vector large $N$ limit, where the boson is a singlet of the $SU(N)$ global symmetry (instead of being in the adjoint), and the fermion is still in the fundamental representation\cite{Polchinski1994, Altshuler1994, Nayak1994a, Lee2009, Mross2010}. The main difference with the matrix large $N$ limit that we have studied here is that now the fermion anomalous dimension is a $1/N$ effect. The resulting RG equations have a similar structure as in \ref{eq:betafcs1} but with a very different $N$ dependence --as a result, the fixed point now occurs for a coupling $\alpha_* \sim \epsilon N$.
Thus, the vector large $N$ limit probes a rather different asymptotic behavior, which is perturbatively accessible only when $\epsilon N \ll 1$.  This is to be contrasted with the matrix large $N$, where the only requirement is that $\epsilon, 1/N \ll 1$ separately, while their ratio $\epsilon N $ can be arbitrary.  Such a regime cannot be controllably accessed in the vector large $N$ model. From the analysis of the RG equations, one finds that for energies $p_0 \ll M_D$, there are two emergent scales, one associated with the development of superconductivity and the other associated with non-Fermi liquid behavior,
\begin{equation}
\mu_{BCS} \sim e^{-\sqrt{N/\alpha}} \Lambda_0, \ \ \ \mu_{NFL} \sim  e^{-1/ \gamma} \Lambda_0  = e^{-\left(\frac{N}{\alpha} \right)} \Lambda_0
\end{equation}
At large $N$, $\mu_{BCS} \gg \mu_{NFL}$ and hence the SC gap is always parametrically larger than the NFL scale. Stated differently, the second limit, $\mu_{BCS} \ll \mu_{NFL}$, requires $ \epsilon \gtrsim 1$, which remains outside the regime where the theory is controllable. This agrees with the analysis of\cite{Metlitski} in a related class of models.

\subsection{Towards a framework for unconventional metals}\label{subsec:unconventional}

Our previous analysis shows that for $\epsilon N>12$ there is no superconducting instability, with $\Delta$ and all its derivatives vanishing continuously. Interestingly, while the superconducting state is extinguished, the Eliashberg equations allow us to determine the state that ensues for $\epsilon N>12$. To see this, note that the 4-Fermi coupling beta function, obtained in (\ref{eq:betalx}), admits a scale invariant fixed point for the BCS coupling
\be\label{eq:lambdaIR}
\lambda_*=\frac{\pi^2}{3} \epsilon N (1-\sqrt{1-12/(\epsilon N)})\,.
\ee
At the same time, since $\Delta =0$, the Eliashberg equation for the fermion self-energy implies that the cubic coupling and anomalous dimension attain fixed point values $\alpha_*=\epsilon/3, \gamma_*=\epsilon/6$. This describes a new state of matter with a ``naked'' quantum critical point, under full perturbative control.
This fixed point is characterized by Cooper pair fluctuations that are critical at all scales.

We propose that this regime can be used as a theoretical model to understand some of the properties of unconventional metals. The main motivation for this is that it provides a framework for studying quantum criticality in a controlled way, with the scale invariant fixed point prevailing over the superconducting instability. While the large $N$ approach, needed to make the analysis controlled, is unlikely to apply directly to any condensed matter system, it is natural to speculate that the solutions obtained this way may still bear resemblance to experiments.  The most robust aspect of our analysis is that the QCP occurs at a {\it finite} value of the BCS coupling.  This is to be contrasted to a Fermi liquid, where the BCS coupling is either zero (for repulsive interactions), or diverges, signaling a superconducting instability for the attractive case.  

Many experimental consequences of such strong superconducting fluctuations are likely to occur in transport and thermodynamic signatures.  For instance, a classic signature of strong superconducting fluctuations is a large, positive magnetoresistance.  Similar transport signatures ought to be present  if the physics of the QCP governs the behavior at scales larger than $T_c$.  An additional consequence of quantum critical superconducting fluctuations is a strong diamagnetic response in the metallic regime above the quantum critical point. Another interesting observable is associated to Friedel oscillations, which originate from singularities with $2k_F$ momentum transfer. Due to the 4-Fermi coupling fixed point, the $2k_F$ vertex receives a large anomalous dimension, while in most models, including Fermi liquid theory, this parameter is negligible.

More broadly, the contributions of such strong, critical superconducting fluctuations to the incoherent metallic behavior in the normal state are largely unexplored, and provide interesting open questions for future work.

\section{Conclusions}\label{sec:conclusion}

In this work we have studied the interplay between quantum criticality and superconductivity for non-Fermi liquids in a general class of models parametrized by an arbitrary 4-Fermi exchange kernel $u(p_0)$. Our main theoretical result is the derivation of the RG beta functions from the Eliashberg equations. The key approximation for this to hold is that quantum renormalizations be approximately local in energy scale, according to the criterion of \S \ref{subsec:local}. Under this condition, we reduced the gap equation to differential form, and constructed an explicit map to the 4-Fermi coupling. The resulting beta function was shown to capture all the expected quantum effects. The condition of energy locality for the gap equation has to be checked for specific models; here we verified it numerically for phonon-mediated superconductivity, and for non-Fermi liquids near their critical dimension. In this way, we established the full equivalence between the Eliashberg and RG predictions. We applied this to the electron-phonon system, and to non-Fermi liquids near symmetry breaking transitions. In both cases, our approach revealed new aspects in their dynamics, such as the gap behavior at moderate couplings in the electron-phonon case, and the BKT transition and quantum critical regime for the NFLs.

Let us end by suggesting directions of future investigation. Our analysis was done at the linearized level and at $T=0$; it will be important to extend this to finite temperature and to include nonlinear effects from the gap. We expect that both extensions will bring in new effects, some of which have already been reported; see\cite{2016arXiv160601252W, GTtalk}. Another theoretical development would be to extend our present analysis to systems where there is no factorization of the momentum integrals in the Eliashberg equations --this is also relevant for working at finite temperature. It is an open question whether for this case the SDE and RG equations are compatible.

Finally, it will be interesting to analyze the phenomenological consequences of this class of models, including transport and finite temperature observables. Even though our analysis was performed at large $N$, we expect that some of the broad features (BKT scaling, critical point with finite BCS scaling) can survive for a small number of flavors. In particular, for the relevant case of two-dimensional systems ($\epsilon=1$), already with a small number of flavors we expect to be close to $N \epsilon \sim 1$ where incoherent effects on SC become important. 

\acknowledgments{We gratefully thank A. Chubukov for many related discussions on non-Fermi liquids.  We thank E. Abrahams, S. Chakravarty, S. Kivelson, and  M. Mulligan for interesting comments on a preliminary version.  SR is supported by the DOE Office of Basic Energy Sciences, contract DE-AC02-76SF00515. GT is supported by CONICET, and PIP grant 11220110100752. HW is supported by DARPA YFA contract D15AP00108.}


\bibliography{NFL}

\end{document}